\documentclass[10pt,american,prl,floatfix,superscriptaddress,twocolumn,showpacs,amsmath,amssymb,reprint]{revtex4-1}
\usepackage[T1]{fontenc}
\usepackage[latin9]{inputenc}
\usepackage{color}
\usepackage{float}
\usepackage{amsmath}
\usepackage{graphicx}

\usepackage{hyperref}
\hypersetup{backref, 
colorlinks=true,
citecolor=blue,
linkcolor=blue,
filecolor=blue,
urlcolor=blue,
breaklinks=true
} 

\makeatletter
\PassOptionsToPackage{caption=false}{subfig} 

\makeatother

\usepackage{babel}
\begin{document}

\title{Emergent SU(3) symmetry in random spin-1 chains}

\author{V. L. Quito}

\affiliation{Instituto de Física Gleb Wataghin, Unicamp, Rua Sérgio Buarque de
Holanda, 777, CEP 13083-859 Campinas, SP, Brazil}

\author{José A. Hoyos}

\affiliation{Instituto de Física de São Carlos, Universidade de São Paulo, C.P.
369, São Carlos, SP 13560-970, Brazil}

\author{E. Miranda}

\affiliation{Instituto de Física Gleb Wataghin, Unicamp, Rua Sérgio Buarque de
Holanda, 777, CEP 13083-859 Campinas, SP, Brazil}

\date{\today}
\begin{abstract}
We show that generic SU(2)-invariant random spin-1 chains have phases
with an emergent SU(3) symmetry. We map out the full zero-temperature
phase diagram and identify two different phases: (i) a conventional
random singlet phase (RSP) of strongly bound spin pairs (SU(3) ``mesons\textquotedbl{})
and (ii) an unconventional RSP of bound SU(3) ``baryons'', which are
formed, in the great majority, by spin trios located at random
positions. The emergent SU(3) symmetry dictates that susceptibilities
and correlation functions of both dipolar and quadrupolar spin operators
have the same asymptotic behavior.
\end{abstract}

\pacs{75.10.Jm, 75.10.Pq, 75.10.Nr}

\maketitle
\emph{Introduction.}---Symmetries constitute a fundamental ingredient
in our description of nature. In the standard model of elementary
particles gauge symmetry is the organizing principle, strongly restricting
the form of the microscopic equations \cite{Cheng1988}. In condensed
matter systems, symmetries play a crucial role in classifying the various
phases and transitions between them \cite{AndersonBook}. Symmetries
can be destroyed at low energies, partially or completely, through
the mechanism of spontaneous symmetry breaking. This is a familiar
theme both in high-energy physics, as exemplified by the electroweak
symmetry breaking, and at the condensed matter scale, in the various
broken-symmetry phases of magnetism, superconductivity, superfluidity
and others. The pattern of symmetry breaking, then, plays an important
role in determining the spectrum of low-energy excitations (Goldstone
bosons, quasiparticles, etc.) in the asymmetric phases \cite{AndersonBook}.

A much less explored phenomenon is the enlargement of a system's symmetry
at low energies. One of the earliest signs of such emergent
symmetries was found in the critical region of the Ising chain in
a transverse field, where the spectrum was predicted to be governed
by the E$_{8}$ Lie group \cite{zamolodchikov89}, which was later confirmed
experimentally \cite{Coldeaetal2010}. Other candidate systems have
been proposed, both at fine-tuned critical points \cite{PhysRevLett.89.277203,Senthietal2004,Groveretal2014}
and in extended phases \cite{linetal98,batistaortiz04,fidkowski-etal-prb09,Yipetal2015}.
Some proposals for quantum simulators of lattice gauge theories using
cold atoms rely on the realization of the gauge symmetry as an emergent
one \cite{zohar2015quantum}.

Although a generic mechanism for the appearance of an emergent symmetry
is not known, it has been suggested that such emergent symmetries arise when the ground state
is a collection of subsystems coupled only by symmetry-breaking terms
that are irrelevant in the renormalization-group (RG) sense \cite{batistaortiz04,schmalianbatista08}.
The emergent symmetry is, then, that of the subsystems.

\begin{figure}[t]
\begin{centering}
\includegraphics[scale=0.25]{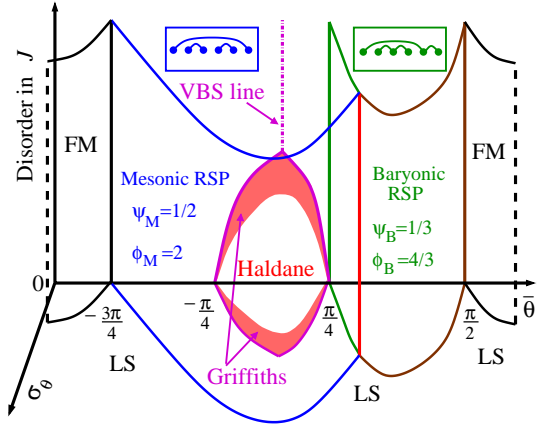}
\par\end{centering}

\raggedright{}\caption{
\label{fig:Phase-diagram}{ 
Phase diagram of the random spin-1 chain. The angle $\theta$ is defined
through the ratio between biquadratic and bilinear couplings $\tan\theta_{i}=\frac{D_{i}}{J_{i}}$.
Solid lines represent continuous transitions. The Haldane phase is
characterized by a finite gap and topological order. In the red (shaded)
Griffiths region the gap vanishes but the topological order remains.
At stronger disorder, there is a conventional ``mesonic''\textcolor{red}{{}
} RSP and an unconventional ``baryonic'' RSP.
In the latter, the singlets are formed mostly out of spin trios. The
insets depict schematically the corresponding random-singlet ground
states. LS and FM stand for Large Spin and ferromagnetic phases, respectively.}
}
\end{figure}

In this Letter, we show that quenched disorder may also lead to an
asymptotically decoupled ground state accompanied by an emergent global
symmetry. We show that the most general disordered antiferromagnetic (AFM)
SU(2)-symmetric spin-1 chain is characterized by an emergent SU(3)
symmetry. At low energies, the system behaves as a collection of decoupled
objects - namely: unbound SU(3) ``quarks'' and ``antiquarks'' plus
bound SU(3) ``mesons'' or ``baryons'' - depending on the phase (see
Fig.~\ref{fig:Phase-diagram}). As a consequence, susceptibilities
and correlation functions of appropriately defined SU(3) operators
(spin dipoles and quadrupoles) are governed by same universal exponents.
Moreover, the emergent symmetry is identified in finite regions
of parameter space and requires no fine tuning. As will become clear,
this mechanism delineates a generic route towards emergent symmetries
in strongly disordered systems.

\emph{The model.}---We consider the most general SU(2)-symmetric random
spin-1 chain given by the Hamiltonian

\begin{eqnarray}
\mathcal{H} & =\sum_{i}\mathcal{H}_{i}= & \sum_{i}\left[J_{i}\mathbf{S}_{i}\cdot\mathbf{S}_{i+1}+D_{i}\left(\mathbf{S}_{i}\cdot\mathbf{S}_{i+1}\right)^{2}\right],\label{eq:hamilt}
\end{eqnarray}
where $J_{i}$ and $D_{i}$ are independent random variables. In addition to
condensed matter realizations, this may be especially relevant in cold-atom
systems where spin couplings, dimensionality, and disorder can be
controlled with considerable flexibility \cite{PhysRevA.68.063602,PhysRevLett.93.250405}. 

The model is analyzed through the strong-disorder renormalization-group
(SDRG) method~\cite{madasguptahu,madasgupta,bhatt-lee} (for a general
review see, e.g., Ref.\ \cite{igloi-review} and for specific spin-1
systems, see \cite{boechatbiquadr,hymanyang97,monthusetal97,monthusetal98,PhysRevLett.89.117202,bergekvist-etal-prb02,PhysRevB.66.104425}),
which gives asymptotically exact results when the effective disorder
grows without bound in the RG sense~\cite{fisher94-xxz}. When this
happens, the system is said to be governed by an infinite randomness
fixed point (IRFP). As we will show, this is the case in the random
singlet phases (RSPs) of our model (see Fig.~\ref{fig:Phase-diagram}).

It is useful to define the variable \textcolor{black}{$\tan\theta_{i}=\frac{D_{i}}{J_{i}}$.}
The clean zero-temperature phase diagram has been extensively studied
and shown to be quite rich (see, e.g., \cite{manmanaetal11}  and
references therein). \textcolor{black}{There is a conventional FM
phase when $\frac{\pi}{2}<\theta<\frac{5\pi}{4}$. The system is gapped
if $-\frac{3\pi}{4}<\theta<\frac{\pi}{4}$ and critical when $\frac{\pi}{4}<\theta<\frac{\pi}{2}$.
For $-\frac{3\pi}{4}\le\theta\le-\frac{\pi}{4}$, the ground state
is spontaneously dimerized. The topological Haldane phase extends
from $-\frac{\pi}{4}$ to $\frac{\pi}{4}$. Moreover, some special
points are noteworthy: the Affleck-Kennedy-Lieb-Tasaki (AKLT) point
($\tan\theta=\frac{1}{3}$, with $J>0$), at which the ground state
is known to be a valence bond solid (VBS), four SU(3)-symmetric points
$\theta=\frac{\pi}{4}$, $\theta=\pm\frac{\pi}{2}$ and $\theta=-\frac{3\pi}{4}$,
and the critical point $\theta=-\frac{\pi}{4}$.}

\emph{The decimation procedure.}---We now describe the SDRG decimation
procedure assuming strong disorder (the weak-disorder regime is discussed
in the Supplemental Material \cite{Suppl}). The idea is to obtain
a description of the low-energy sector by gradually eliminating high-energy
excitations of small clusters and finding the effective Hamiltonian
of the remaining degrees of freedom. We define the $i$th gap $\Delta_{i}$
as the energy difference between the ground and the first excited
state of the local Hamiltonian ${\cal H}_{i}$. At each step, we look
for the largest gap say, $\Delta_{2}=\max\left(\Delta_{i}\right)\equiv\Omega$,
keep only the lowest-energy multiplet of ${\cal H}_{2}$, and use
perturbation theory to find how the remaining degrees of freedom are
coupled. 

\begin{figure}[b]
\begin{centering}
\includegraphics[clip,scale=0.14]{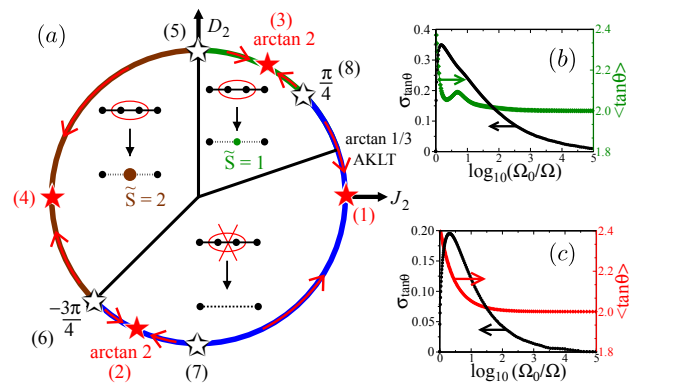}
\par\end{centering}

\caption{\label{fig:decimation} (a) SDRG decimation steps,
fixed points and basins of attraction. Solid (open) stars denote stable
(unstable) fixed points. (b) and (c) show the average and width of the $\tan\theta_{i}$
distribution along the SDRG flow. Initially, (b) $\theta_{i}=3\pi/8$
and (c) $\theta_{i}=-5\pi/8$, with $\sigma_{\tan\theta}=0$ in both
cases.}
\end{figure}

The possible steps are depicted in Fig.~\ref{fig:decimation}(a).
When the ground state of ${\cal H}_{2}$ is a singlet ($-\frac{3\pi}{4}<\theta_{2}<\arctan\frac{1}{3}$),
spins $S_{2}$ and $S_{3}$ are removed and the new effective couplings
between spins $S_{1}$ and $S_{4}$ are
\begin{eqnarray}
\tilde{K} & = & \frac{4K_{1}K_{3}}{3\left(K_{2}-\frac{5}{2}D_{2}\right)},\,\,\,\tilde{D}=-\frac{2D_{1}D_{3}}{9\left(K_{2}-\frac{1}{2}D_{2}\right)},\label{eq:KD-2nd}
\end{eqnarray}
where $K_{i}=J_{i}-D_{i}/2$ \footnote{$K_{i}$ is a natural coupling constant when \eqref{eq:hamilt} is
written in terms of irreducible spherical tensors, as noted in \cite{PhysRevLett.80.4562}}. On the other hand, when the ground state is a triplet ($\arctan\frac{1}{3}<\theta_{2}<\frac{\pi}{2}$),
the pair is replaced by a new spin 1 degree of freedom coupled to $S_{1}$
and $S_{4}$ via 
\begin{eqnarray}
\tilde{K}_{i} & = & \frac{1}{2}K_{i},\,\,\,\tilde{D}_{i}=-\frac{1}{2}D_{i},\label{eq:KD-1st}
\end{eqnarray}
with $i=1,3$. Finally, when $\frac{\pi}{2}<\theta_{2}<\frac{5\pi}{4}$,
the ground state is a quintuplet, the spin pair can be replaced by
an effective spin-2 degree of freedom and the new effective couplings
are $\tilde{K}_{i}=\frac{1}{2}K_{i}$ and $\tilde{D}_{i}=\frac{1}{24}D_{i}$.
It turns out that the renormalized Hamiltonian retains the form~\eqref{eq:hamilt},
albeit with different spins and new couplings, as previously reported
in~\cite{PhysRevLett.80.4562}. 

\emph{SDRG flows}.---Inspection of Eqs.~(\ref{eq:KD-2nd}) and (\ref{eq:KD-1st})
allows us to immediately identify four fixed points (FPs) of
the SDRG flow. They are characterized by fixed angles $\theta_{i}$.
We denote them by numbers \textcolor{black}{{[}see solid red stars
in Fig.~\ref{fig:decimation}(a){]}, }as follows.

(1) The FP $D_{i}=0$, with $K_{i}=J_{i}>0$ ($\theta_{i}=0$),
is the disordered AFM Heisenberg chain, which was intensively studied in Refs. 
\cite{boechatbiquadr,hymanyang97,monthusetal97,monthusetal98,PhysRevLett.89.117202}.
For strong enough disorder, the flow is towards an IRFP (the relative
width of the distribution of couplings grows without bounds) and the
ground state is a collection of nearly independent singlets formed
between arbitrarily distant spin pairs - a conventional random singlet
(RS) state.

(2) The FP $K_{i}=0$ with $D_{i}<0$ ($\theta_{i}=\arctan2$ in the
third quadrant), which corresponds to a flow similar to FP (1) since
all decimations are of singlet-formation type [see Fig.~\ref{fig:decimation}(a)]
and lead to the same conventional RS state.

(3) The FP $K_{i}=0$ with positive and negative $D_{i}$ \textcolor{black}{($\theta_{i}=\arctan2$,
with $\theta_{i}$ in both the first and third quadrants)}. This FP involves
SDRG steps of both singlet- and triplet-generating types [see Fig.~\ref{fig:decimation}(a)].
Here, the effective system has equal fractions of positive and
negative $\tilde{D}_{i}$'s (and, hence, equal fractions of singlet-
and triplet-generating decimations), since this is the only situation
that is preserved by the RG flow. Although this FP actually encompasses
both signs of $D_{i}$, we choose to represent it by a single star
in the first quadrant. The presence of both types of decimations leads
to a state different from FPs (1) and (2) above. In fact, this SDRG
flow is, up to irrelevant numerical prefactors, identical to that
of generic SU(3)-symmetric chains discussed in \cite{PhysRevB.70.180401}.
We will come back to this point later. 

(4) The FP $D_{i}=0$, with $K_{i}=J_{i}<0$ ($\theta_{i}=\pi$),
which corresponds to the disordered FM Heisenberg chain. \textcolor{black}{As
spins larger than 2 are generated in this case, the decimation procedure
must be complemented by those of ref.~\cite{westerbergetal}. This
FM state is not the focus of this Letter and will be considered elsewhere~\cite{quito-hoyos-miranda}}.

It is straightforward to show \textcolor{black}{\cite{Suppl}} that
the FPs (1)-(3) are stable with respect to narrow distributions of angles
$\theta_{i}$\textcolor{black}{. Moreover, our extensive numerics
indeed confirm that these are the only four stable FPs. Thus, there
must be four other unstable ones {[}depicted as open stars in }Fig.~\ref{fig:decimation}(a)\textcolor{black}{{]}.
}The strongest candidates are the four SU(3)-symmetric points $\theta_{i}=\pm\frac{\pi}{2},\frac{\pi}{4},$
and $-\frac{3\pi}{4}$, since this global symmetry is preserved by
the SDRG. Although our methods cannot be used precisely \emph{at}
the FM points $\theta_{i}=\frac{\pi}{2}$ and $\theta_{i}=-\frac{3\pi}{4}$,
we have checked numerically that the \textcolor{black}{SD}RG flow
is always away from them \cite{Suppl}. We thus conjecture they are
unstable FM FPs and denote them by (5) and (6), respectively.

\textcolor{black}{We now show that the other two, $\theta_{i}=\frac{\pi}{4}$
and $\theta_{i}=-\frac{\pi}{2}$, are indeed FPs.} \textcolor{black}{As
it will be important for the understanding of the emergent SU(3) symmetry,
we will describe these points in detail. The Hamiltonian at these
points can be recast as \cite{Suppl}} 
\begin{equation}
H=\sum_{i}\sum_{a=1}^{8}C_{i}\Lambda_{a,i}\cdot\Lambda_{a,i+1}+\mathrm{const}\label{eq:mesonicchain}
\end{equation}
where $\Lambda_{a,i}$ ($a=1,\ldots,8$) are the generators of an
irreducible representation (IR) of SU(3). When $\theta_{i}=-\frac{\pi}{2}$,
$C_{i}=\left|D_{i}\right|/2$ and the IR on odd (even) sites is the
fundamental (antifundamental) one, also called the quark (antiquark) IR.
As can be verified from Eq.~\eqref{eq:KD-2nd}, this FP is characterized
by singlet formation only. At each step a quark binds to an antiquark
to form a singlet (a meson, in QCD language). Note that the alternation
of quarks and antiquarks in the chain is preserved by this flow. It
thus realizes the same kind of RS state of the FPs (1) and (2) above.
We will, accordingly, dub it a mesonic RS state and number it as (7). 

When $\theta_{i}=\frac{\pi}{4}$, the IR is the quark one on every
site and $C_{i}=J_{i}/2$. Decimation of a bond with $\frac{\pi}{4}$
turns the adjacent bond angles into $-\frac{\pi}{2}$ {[}see\textcolor{black}{{}
Eq.~(\ref{eq:KD-1st})}{]}. There is, thus, a quick proliferation of
bonds with $\theta_{i}=-\frac{\pi}{2}$. \textcolor{black}{Further
decimation of a bond with $\theta_{i}=-\frac{\pi}{2}$ leads to a
spin singlet and an effective bond angle of $-\frac{\pi}{2}$, if
$\theta_{i-1}=\theta_{i+1}$, or $\frac{\pi}{4}$ otherwise {[}see
Eqs.~(\ref{eq:KD-2nd}) and (\ref{eq:KD-1st}){]}.} The \textcolor{black}{SDRG
flow for this} FP, which we will number as (8), is characterized by
 equal fractions of bonds with $\theta_{i}=\frac{\pi}{4}$ and
$\theta_{i}=-\frac{\pi}{2}$ (again, this is the only situation preserved
by the flow \cite{Suppl}). In SU(3) language, two original quarks
first bind to form an effective antiquark \cite{PhysRevB.70.180401}
{[}the effective spin 1 of the decimation step of Eq.~\eqref{eq:KD-1st}{]}.
This antiquark can later bind to a third quark to form a singlet.
Effectively, this singlet is formed out of three original quarks,
just like a baryon is formed out of three valence quarks \footnote{In general, singlets of 6, 9,... original spins/quarks are also formed,
though less abundantly \cite{PhysRevB.70.180401}.}. Note that the structure of this SDRG flow is the same as that
of FP (3), even though the couplings and angles are not the same. We
call this a baryonic RS state. For simplicity, as in case (3), this
case (8) is represented somewhat imprecisely by a single star in the
first quadrant.

\emph{Phase diagram}.---The identification of all FPs and their stability
properties allow for the immediate description of three basins of attraction
for initial strongly disordered distributions of coupling constants
with a fixed angle $\theta_{i}=\theta_{0}$ for all
$i$ (the initial angle distribution width $\sigma_{\theta}=0$),
as shown by the differently colored arcs along the circumference of
Fig.~\ref{fig:decimation}(a). The red arrows show the flow direction,
with the caveat that the flow towards the FP (3) (the green arc) also
involves excursions into region $-\frac{3\pi}{4}<\theta\le-\frac{\pi}{2}$.
We have \textcolor{black}{verified} numerically that these are indeed
the only possible flows \cite{Suppl}. Two typical examples are shown
in Figs.~\ref{fig:decimation}(b) and (c). Note that the distribution
widths initially grow but eventually vanish as the stable FPs are
approached. This is to be expected, since the stable FPs are characterized
by a unique value of $\tan\theta$. In other words, $\sigma_{\tan\theta}$
is an irrelevant variable at the stable FPs. As discussed in
greater detail in the Supplemental Material \cite{Suppl}, we only
expect the behavior found in the strong-disorder regime to break down
inside a dome around the Haldane phase $-\frac{\pi}{4}<\theta_{0}<\frac{\pi}{4}$.
This allows us to obtain the phase diagram shown in Fig.~\ref{fig:Phase-diagram}\textcolor{black}{{}
on the plane $\sigma_{\theta}=0$.}

We now describe the physical properties of the various phases. In
the whole region $-\frac{3\pi}{4}<\theta_{0}<\frac{\pi}{4}$ {[}the
blue arc of Fig.\ \ref{fig:decimation}(a){]}, all decimations lead,
after an initial transient, to the formation of ever-more-widely separated
singlet pairs (no trios) and the ground state is analogous to the
RSP of the spin-1/2 AFM Heisenberg chain~\cite{fisher94-xxz}. The
flow is attracted by either of the two stable FPs (1) and (2). Since
their structure is the same as the unstable SU(3)-symmetric FP (7),
we describe this whole region as a mesonic RSP. The properties of
such phases are well known~\cite{fisher94-xxz}. The energy ($\Omega$)
and length ($L$) scales of excitations obey activated dynamical scaling
$\ln\Omega\sim-L^{\psi}$ with a universal $\psi=\psi_{M}=\frac{1}{2}$,
the magnetic susceptibility diverges as $\chi\sim1/\left(T\left|\ln T\right|^{1/\psi}\right)$,
and the specific heat vanishes as $c\sim\left|\ln T\right|^{-\left(1+1/\psi\right)}$
as $T\rightarrow0$. The typical ground-state spin-spin correlations
vanish as $\sim\exp\left(-{\rm const}\times\left|i-j\right|^{\psi}\right)$,
as a consequence of the localized nature of the phase, whereas the
average correlations are dominated by the spin singlets and vanish
only algebraically $\sim e^{iq\left(i-j\right)}\left|i-j\right|^{-\phi}$,
with $q=q_{M}=\pi$ and a universal exponent $\phi=\phi_{M}=2$. The
difference between the two FPs lies in the nature of the excitations.
\textcolor{black}{For }$-\frac{\pi}{2}<\theta_{0}<\frac{\pi}{4}$,
the lowest excitation of a random singlet pair has spin 1, whereas
\textcolor{black}{for }$-\frac{3\pi}{4}<\theta_{0}<-\frac{\pi}{2}$,
it has spin 2. At the SU(3)-symmetric FP (7), the two types of excitations
become degenerate and are analogous to the meson octuplets of
QCD.

In the region $\frac{\pi}{4}<\theta_{0}<\frac{\pi}{2}$ {[}the green
arc of Fig.~\ref{fig:decimation}(a){]} the flow converges to the
FP (3) and is characterized by the formation of baryonic-like singlet
trios (and also rarer sextets, etc.). In fact, the same happens for
$\theta_{0}=\frac{\pi}{4}$ {[}the unstable SU(3)-symmetric one FP
(8){]}. As a result, as shown in \cite{PhysRevB.70.180401}, the low-energy
physical properties of this baryonic RSP have the same generic forms
\textcolor{black}{as in the mesonic RSP}, but with $q=q_{B}=2\pi/3$
and the important difference that the universal exponents change
to $\psi=\psi_{B}=1/3$ and $\phi=\phi_{B}=4/3$.

We now address the case when the initial distribution of angles has
a nonzero width\textcolor{black}{{} $\sigma_{\theta}$}. We have verified
numerically that the phase diagram is still valid as long as \emph{all}
the initial angles lie inside the basin of attraction of the corresponding
phase. Otherwise, the flow is more involved. When the mesonic
and the baryonic RSPs initially compete, the former absorbs the flow. When the
FM phase competes with any of the others, the system flows \textcolor{black}{to
the so-called Large Spin phase~\cite{westerbergetal}, as a consequence
of the presence of both AFM and FM couplings.}\textcolor{red}{{} }With
this, we complete the topology of the phase diagram of Fig.~\ref{fig:Phase-diagram}.

\emph{Emergent SU(3) }\textcolor{black}{\emph{symmetry}}.---A few
elements of the flow to an IRFP completely determine the low- and zero-temperature properties of the system. In the present case, these elements
are (i) a ground state made up of strongly coupled singlet-forming
spins, pairs in the case of FPs (1), (2), and (7), and (mostly) trios
at FPs (3) and (8), with well-characterized size distributions \cite{fisher94-xxz,PhysRevB.70.180401,PhysRevB.76.174425},
and (ii) low-energy excitations consisting of essentially free spin-1
clusters with known scale-dependent density \cite{bhatt-lee,fisher94-xxz,PhysRevB.70.180401}.
We now show that both elements have SU(3) symmetry and this gives
rise to an emergent SU(3) symmetry in extended regions of the
phase diagram. Indeed, the singlets of element (i) are not only SU(2)
but also SU(3) singlets. In addition, the spin-1 clusters of element (ii)
transform as SU(3) quarks or antiquarks. This can be more easily seen
from the fact that the ground multiplets of the singlet- and triplet-generating
decimation steps of Eqs.~(\ref{eq:KD-2nd}) and (\ref{eq:KD-1st})
are all $\theta$-independent. In other words, the ground states
at the FPs (1), (2) and (7) are the same, and so are the ground states
of FPs (3) and (8). This leads immediately to the result that, at
$T=0$, the average and typical correlation functions of all
SU(3) generators $\Lambda_{a}\,\left(a=1,\ldots8\right)$, which
include dipolar and quadrupolar spin operators, are governed by the
same exponents $\left(\psi_{H},\phi_{H}\right)$ with $H=M$ or $B$.
For a complete list of these operators, see the Supplemental Material
\cite{Suppl}.

Likewise, the low-temperature SU(3) susceptibilities of the RSPs can
be written as \cite{bhatt-lee,PhysRevB.70.180401} 
\begin{equation}
\chi_{a}\left(T\right)\approx n\left(\Omega=T\right)\chi_{a}^{\text{free}}\left(T\right),
\end{equation}
where $n\left(\Omega\right)=\frac{N\left(\Omega\right)}{L_{0}}$ is
the density of undecimated spin clusters at the scale $\Omega$ and
$\chi_{a}^{\text{free}}\left(T\right)$ is the SU(3) susceptibility of a
free spin cluster. Since the free spin clusters (triplets) are SU(3)
quarks or antiquarks, $\chi_{a}^{\text{free}}\left(T\right)$ is SU(3) symmetric
and independent of $a$. Taking, e.g., $\Lambda_{3}=\tilde{\Lambda}_{3}=S_{z}$,
$\chi_{a}^{\text{free}}\left(T\right)=2/\left(3T\right)$ for both representations.
Since $n\left(\Omega=T\right)\sim1/\left|\ln T\right|^{1/\psi_{H}}$
\cite{Suppl}, we get $\chi_{a}\sim1/\left(T\left|\ln T\right|^{1/\psi_{H}}\right)$. 

The emergent symmetry occurs even at the Heisenberg point (1), a feature
previously unnoticed. We stress that although these various quantities
are all governed by the same exponents, the numerical prefactors are
\emph{not} the same due to the initial inexactness of the SDRG procedure.
A similar phenomenon is observed in disordered spin-$\frac{1}{2}$
XXZ chains, in which, despite the absence of global SU(2) symmetry,
both longitudinal and \textcolor{black}{transverse correlations and
susceptibilities are governed by the same exponents \cite{fisher94-xxz}.}

\emph{Conclusions.}--- We have found the generic route towards emergent
symmetries at IRFPs: the ground multiplets of the two-spin problem (of
the SDRG) must transform as irreducible representations of the emergent
symmetry. In the present case, the singlet and the triplet states
are SU(3) symmetric. The simplicity of the mechanism responsible for
the emergent symmetry that we uncovered suggests that it might find other
realizations. We have looked for them in generic disordered
SU(2)-invariant spin-$S$ chains with $S>1$ and, surprisingly, found none, although
we did find cases with $\psi\neq\frac{1}{2}$ \textcolor{black}{\cite{quito-hoyos-miranda}}.
Thus, as in the other known realizations of emergent symmetries, finding
a recipe for generating them poses a problem that remains wide open.

\emph{Acknowledgments}--- We would like to acknowledge financial support from FAPESP and CNPq.

\bibliographystyle{apsrev4-1}

\begin{thebibliography}{36}%
\makeatletter
\providecommand \@ifxundefined [1]{%
 \@ifx{#1\undefined}
}%
\providecommand \@ifnum [1]{%
 \ifnum #1\expandafter \@firstoftwo
 \else \expandafter \@secondoftwo
 \fi
}%
\providecommand \@ifx [1]{%
 \ifx #1\expandafter \@firstoftwo
 \else \expandafter \@secondoftwo
 \fi
}%
\providecommand \natexlab [1]{#1}%
\providecommand \enquote  [1]{``#1''}%
\providecommand \bibnamefont  [1]{#1}%
\providecommand \bibfnamefont [1]{#1}%
\providecommand \citenamefont [1]{#1}%
\providecommand \href@noop [0]{\@secondoftwo}%
\providecommand \href [0]{\begingroup \@sanitize@url \@href}%
\providecommand \@href[1]{\@@startlink{#1}\@@href}%
\providecommand \@@href[1]{\endgroup#1\@@endlink}%
\providecommand \@sanitize@url [0]{\catcode `\\12\catcode `\$12\catcode
  `\&12\catcode `\#12\catcode `\^12\catcode `\_12\catcode `\%12\relax}%
\providecommand \@@startlink[1]{}%
\providecommand \@@endlink[0]{}%
\providecommand \url  [0]{\begingroup\@sanitize@url \@url }%
\providecommand \@url [1]{\endgroup\@href {#1}{\urlprefix }}%
\providecommand \urlprefix  [0]{URL }%
\providecommand \Eprint [0]{\href }%
\providecommand \doibase [0]{http://dx.doi.org/}%
\providecommand \selectlanguage [0]{\@gobble}%
\providecommand \bibinfo  [0]{\@secondoftwo}%
\providecommand \bibfield  [0]{\@secondoftwo}%
\providecommand \translation [1]{[#1]}%
\providecommand \BibitemOpen [0]{}%
\providecommand \bibitemStop [0]{}%
\providecommand \bibitemNoStop [0]{.\EOS\space}%
\providecommand \EOS [0]{\spacefactor3000\relax}%
\providecommand \BibitemShut  [1]{\csname bibitem#1\endcsname}%
\let\auto@bib@innerbib\@empty
\bibitem [{\citenamefont {Cheng}\ and\ \citenamefont {Li}(1988)}]{Cheng1988}%
  \BibitemOpen
  \bibfield  {author} {\bibinfo {author} {\bibfnamefont {T.-P.}\ \bibnamefont
  {Cheng}}\ and\ \bibinfo {author} {\bibfnamefont {L.-F.}\ \bibnamefont {Li}},\
  }\enquote {\bibinfo {title} {Gauge {T}heory of {E}lementary {P}article
  {P}hysics},}\ \ (\bibinfo {address} {Oxford},\ \bibinfo {year}
  {1988})\BibitemShut {NoStop}
\bibitem [{\citenamefont {Anderson}(1984)}]{AndersonBook}%
  \BibitemOpen
  \bibfield  {author} {\bibinfo {author} {\bibfnamefont {P.~W.}\ \bibnamefont
  {Anderson}},\ }\enquote {\bibinfo {title} {Basic notions of condensed matter
  physics},}\ \ (\bibinfo  {publisher} {Benjamin-Cummings, Melon-Park, California},\ \bibinfo {year}
  {1984})\BibitemShut {NoStop}%
\bibitem [{\citenamefont {Zamolodchikov}(1989)}]{zamolodchikov89}%
  \BibitemOpen
  \bibfield  {author} {\bibinfo {author} {\bibfnamefont {A.~B.}\ \bibnamefont
  {Zamolodchikov}},\ }\href@noop {} {\bibfield  {journal} {\bibinfo  {journal}
  {Int. J. Mod. Phys. A}\ }\textbf {\bibinfo {volume} {4}},\ \bibinfo {pages}
  {4235} (\bibinfo {year} {1989})}\BibitemShut {NoStop}%
\bibitem [{\citenamefont {Coldea}\ \emph {et~al.}(2010)\citenamefont {Coldea},
  \citenamefont {Tennant}, \citenamefont {Wheeler}, \citenamefont {Wawrzynska},
  \citenamefont {Prabhakaran}, \citenamefont {Telling}, \citenamefont
  {Habicht}, \citenamefont {Smeibidl},\ and\ \citenamefont
  {Kiefer}}]{Coldeaetal2010}%
  \BibitemOpen
  \bibfield  {author} {\bibinfo {author} {\bibfnamefont {R.}~\bibnamefont
  {Coldea}}, \bibinfo {author} {\bibfnamefont {D.~A.}\ \bibnamefont {Tennant}},
  \bibinfo {author} {\bibfnamefont {E.~M.}\ \bibnamefont {Wheeler}}, \bibinfo
  {author} {\bibfnamefont {E.}~\bibnamefont {Wawrzynska}}, \bibinfo {author}
  {\bibfnamefont {D.}~\bibnamefont {Prabhakaran}}, \bibinfo {author}
  {\bibfnamefont {M.}~\bibnamefont {Telling}}, \bibinfo {author} {\bibfnamefont
  {K.}~\bibnamefont {Habicht}}, \bibinfo {author} {\bibfnamefont
  {P.}~\bibnamefont {Smeibidl}}, \ and\ \bibinfo {author} {\bibfnamefont
  {K.}~\bibnamefont {Kiefer}},\ }\href {\doibase 10.1126/science.1180085}
  {\bibfield  {journal} {\bibinfo  {journal} {Science}\ }\textbf {\bibinfo
  {volume} {327}},\ \bibinfo {pages} {177} (\bibinfo {year}
  {2010})}\BibitemShut {NoStop}%
\bibitem [{\citenamefont {Damle}\ and\ \citenamefont
  {Huse}(2002)}]{PhysRevLett.89.277203}%
  \BibitemOpen
  \bibfield  {author} {\bibinfo {author} {\bibfnamefont {K.}~\bibnamefont
  {Damle}}\ and\ \bibinfo {author} {\bibfnamefont {D.~A.}\ \bibnamefont
  {Huse}},\ }\href {\doibase 10.1103/PhysRevLett.89.277203} {\bibfield
  {journal} {\bibinfo  {journal} {Phys. Rev. Lett.}\ }\textbf {\bibinfo
  {volume} {89}},\ \bibinfo {pages} {277203} (\bibinfo {year}
  {2002})}\BibitemShut {NoStop}%
\bibitem [{\citenamefont {Senthil}\ \emph {et~al.}(2004)\citenamefont
  {Senthil}, \citenamefont {Vishwanath}, \citenamefont {Balents}, \citenamefont
  {Sachdev},\ and\ \citenamefont {Fisher}}]{Senthietal2004}%
  \BibitemOpen
  \bibfield  {author} {\bibinfo {author} {\bibfnamefont {T.}~\bibnamefont
  {Senthil}}, \bibinfo {author} {\bibfnamefont {A.}~\bibnamefont {Vishwanath}},
  \bibinfo {author} {\bibfnamefont {L.}~\bibnamefont {Balents}}, \bibinfo
  {author} {\bibfnamefont {S.}~\bibnamefont {Sachdev}}, \ and\ \bibinfo
  {author} {\bibfnamefont {M.~P.~A.}\ \bibnamefont {Fisher}},\ }\href {\doibase
  10.1126/science.1091806} {\bibfield  {journal} {\bibinfo  {journal}
  {Science}\ }\textbf {\bibinfo {volume} {303}},\ \bibinfo {pages} {1490}
  (\bibinfo {year} {2004})}\BibitemShut {NoStop}%
\bibitem [{\citenamefont {Grover}\ \emph {et~al.}(2014)\citenamefont {Grover},
  \citenamefont {Sheng},\ and\ \citenamefont {Vishwanath}}]{Groveretal2014}%
  \BibitemOpen
  \bibfield  {author} {\bibinfo {author} {\bibfnamefont {T.}~\bibnamefont
  {Grover}}, \bibinfo {author} {\bibfnamefont {D.~N.}\ \bibnamefont {Sheng}}, \
  and\ \bibinfo {author} {\bibfnamefont {A.}~\bibnamefont {Vishwanath}},\
  }\href {\doibase 10.1126/science.1248253} {\bibfield  {journal} {\bibinfo
  {journal} {Science}\ }\textbf {\bibinfo {volume} {344}},\ \bibinfo {pages}
  {280} (\bibinfo {year} {2014})}\BibitemShut {NoStop}%
\bibitem [{\citenamefont {Lin}\ \emph {et~al.}(1998)\citenamefont {Lin},
  \citenamefont {Balents},\ and\ \citenamefont {Fisher}}]{linetal98}%
  \BibitemOpen
  \bibfield  {author} {\bibinfo {author} {\bibfnamefont {H.-H.}\ \bibnamefont
  {Lin}}, \bibinfo {author} {\bibfnamefont {L.}~\bibnamefont {Balents}}, \ and\
  \bibinfo {author} {\bibfnamefont {M.~P.~A.}\ \bibnamefont {Fisher}},\ }\href
  {\doibase 10.1103/PhysRevB.58.1794} {\bibfield  {journal} {\bibinfo
  {journal} {Phys. Rev. B}\ }\textbf {\bibinfo {volume} {58}},\ \bibinfo
  {pages} {1794} (\bibinfo {year} {1998})}\BibitemShut {NoStop}%
\bibitem [{\citenamefont {Batista}\ and\ \citenamefont
  {Ortiz}(2004)}]{batistaortiz04}%
  \BibitemOpen
  \bibfield  {author} {\bibinfo {author} {\bibfnamefont {C.~D.}\ \bibnamefont
  {Batista}}\ and\ \bibinfo {author} {\bibfnamefont {G.}~\bibnamefont
  {Ortiz}},\ }\href {\doibase 10.1080/00018730310001642086} {\bibfield
  {journal} {\bibinfo  {journal} {Adv. Phys.}\ }\textbf {\bibinfo {volume}
  {53}},\ \bibinfo {pages} {1} (\bibinfo {year} {2004})}\BibitemShut {NoStop}%
\bibitem [{\citenamefont {Fidkowski}\ \emph {et~al.}(2009)\citenamefont
  {Fidkowski}, \citenamefont {Lin}, \citenamefont {Titum},\ and\ \citenamefont
  {Refael}}]{fidkowski-etal-prb09}%
  \BibitemOpen
  \bibfield  {author} {\bibinfo {author} {\bibfnamefont {L.}~\bibnamefont
  {Fidkowski}}, \bibinfo {author} {\bibfnamefont {H.-H.}\ \bibnamefont {Lin}},
  \bibinfo {author} {\bibfnamefont {P.}~\bibnamefont {Titum}}, \ and\ \bibinfo
  {author} {\bibfnamefont {G.}~\bibnamefont {Refael}},\ }\href {\doibase
  10.1103/PhysRevB.79.155120} {\bibfield  {journal} {\bibinfo  {journal} {Phys.
  Rev. B}\ }\textbf {\bibinfo {volume} {79}},\ \bibinfo {pages} {155120}
  (\bibinfo {year} {2009})}\BibitemShut {NoStop}%
\bibitem [{\citenamefont {Chen}\ \emph {et~al.}(2015)\citenamefont {Chen},
  \citenamefont {Xue}, \citenamefont {McCulloch}, \citenamefont {Chung},
  \citenamefont {Huang},\ and\ \citenamefont {Yip}}]{Yipetal2015}%
  \BibitemOpen
  \bibfield  {author} {\bibinfo {author} {\bibfnamefont {P.}~\bibnamefont
  {Chen}}, \bibinfo {author} {\bibfnamefont {Z.-L.}\ \bibnamefont {Xue}},
  \bibinfo {author} {\bibfnamefont {I.~P.}\ \bibnamefont {McCulloch}}, \bibinfo
  {author} {\bibfnamefont {M.-C.}\ \bibnamefont {Chung}}, \bibinfo {author}
  {\bibfnamefont {C.-C.}\ \bibnamefont {Huang}}, \ and\ \bibinfo {author}
  {\bibfnamefont {S.-K.}\ \bibnamefont {Yip}},\ }\href {\doibase
  10.1103/PhysRevLett.114.145301} {\bibfield  {journal} {\bibinfo  {journal}
  {Phys. Rev. Lett.}\ }\textbf {\bibinfo {volume} {114}},\ \bibinfo {pages}
  {145301} (\bibinfo {year} {2015})}\BibitemShut {NoStop}%
\bibitem [{\citenamefont {Zohar}\ \emph {et~al.}(2015)\citenamefont {Zohar},
  \citenamefont {Cirac},\ and\ \citenamefont {Reznik}}]{zohar2015quantum}%
  \BibitemOpen
  \bibfield  {author} {\bibinfo {author} {\bibfnamefont {E.}~\bibnamefont
  {Zohar}}, \bibinfo {author} {\bibfnamefont {J.~I.}\ \bibnamefont {Cirac}}, \
  and\ \bibinfo {author} {\bibfnamefont {B.}~\bibnamefont {Reznik}},\
  }\href@noop {} {\bibfield  {journal} {\bibinfo  {journal} {arXiv:1503.02312}\
  }}\BibitemShut {NoStop}%
\bibitem [{\citenamefont {Schmalian}\ and\ \citenamefont
  {Batista}(2008)}]{schmalianbatista08}%
  \BibitemOpen
  \bibfield  {author} {\bibinfo {author} {\bibfnamefont {J.}~\bibnamefont
  {Schmalian}}\ and\ \bibinfo {author} {\bibfnamefont {C. D.}~\bibnamefont
  {Batista}},\ }\href {\doibase 10.1103/PhysRevB.77.094406} {\bibfield
  {journal} {\bibinfo  {journal} {Phys. Rev. B}\ }\textbf {\bibinfo {volume}
  {77}},\ \bibinfo {pages} {094406} (\bibinfo {year} {2008})}\BibitemShut
  {NoStop}%
\bibitem [{\citenamefont {Imambekov}\ \emph {et~al.}(2003)\citenamefont
  {Imambekov}, \citenamefont {Lukin},\ and\ \citenamefont
  {Demler}}]{PhysRevA.68.063602}%
  \BibitemOpen
  \bibfield  {author} {\bibinfo {author} {\bibfnamefont {A.}~\bibnamefont
  {Imambekov}}, \bibinfo {author} {\bibfnamefont {M.}~\bibnamefont {Lukin}}, \
  and\ \bibinfo {author} {\bibfnamefont {E.}~\bibnamefont {Demler}},\ }\href
  {\doibase 10.1103/PhysRevA.68.063602} {\bibfield  {journal} {\bibinfo
  {journal} {Phys. Rev. A}\ }\textbf {\bibinfo {volume} {68}},\ \bibinfo
  {pages} {063602} (\bibinfo {year} {2003})}\BibitemShut {NoStop}%
\bibitem [{\citenamefont {Garc\'{\i}a-Ripoll}\ \emph
  {et~al.}(2004)\citenamefont {Garc\'{\i}a-Ripoll}, \citenamefont
  {Martin-Delgado},\ and\ \citenamefont {Cirac}}]{PhysRevLett.93.250405}%
  \BibitemOpen
  \bibfield  {author} {\bibinfo {author} {\bibfnamefont {J.~J.}\ \bibnamefont
  {Garc\'{\i}a-Ripoll}}, \bibinfo {author} {\bibfnamefont {M.~A.}\ \bibnamefont
  {Martin-Delgado}}, \ and\ \bibinfo {author} {\bibfnamefont {J.~I.}\
  \bibnamefont {Cirac}},\ }\href {\doibase 10.1103/PhysRevLett.93.250405}
  {\bibfield  {journal} {\bibinfo  {journal} {Phys. Rev. Lett.}\ }\textbf
  {\bibinfo {volume} {93}},\ \bibinfo {pages} {250405} (\bibinfo {year}
  {2004})}\BibitemShut {NoStop}%
\bibitem [{\citenamefont {Ma}\ \emph {et~al.}(1979)\citenamefont {Ma},
  \citenamefont {Dasgupta},\ and\ \citenamefont {Hu}}]{madasguptahu}%
  \BibitemOpen
  \bibfield  {author} {\bibinfo {author} {\bibfnamefont {S.~K.}\ \bibnamefont
  {Ma}}, \bibinfo {author} {\bibfnamefont {C.}~\bibnamefont {Dasgupta}}, \ and\
  \bibinfo {author} {\bibfnamefont {C.~K.}\ \bibnamefont {Hu}},\ }\href
  {\doibase 10.1103/PhysRevLett.43.1434} {\bibfield  {journal} {\bibinfo
  {journal} {Phys. Rev. Lett.}\ }\textbf {\bibinfo {volume} {43}},\ \bibinfo
  {pages} {1434} (\bibinfo {year} {1979})}\BibitemShut {NoStop}%
\bibitem [{\citenamefont {{C. Dasgupta}}\ and\ \citenamefont {{S.-k.
  Ma}}(1980)}]{madasgupta}%
  \BibitemOpen
  \bibfield  {author} {\bibinfo {author} {\bibnamefont {{C. Dasgupta}}}\ and\
  \bibinfo {author} {\bibnamefont {{S. K. Ma}}},\ }\href {\doibase
  10.1103/PhysRevB.22.1305} {\bibfield  {journal} {\bibinfo  {journal} {Phys.
  Rev. B}\ }\textbf {\bibinfo {volume} {22}},\ \bibinfo {pages} {1305}
  (\bibinfo {year} {1980})}\BibitemShut {NoStop}%
\bibitem [{\citenamefont {Bhatt}\ and\ \citenamefont {Lee}(1982)}]{bhatt-lee}%
  \BibitemOpen
  \bibfield  {author} {\bibinfo {author} {\bibfnamefont {R.~N.}\ \bibnamefont
  {Bhatt}}\ and\ \bibinfo {author} {\bibfnamefont {P.~A.}\ \bibnamefont
  {Lee}},\ }\href {\doibase 10.1103/PhysRevLett.48.344} {\bibfield  {journal}
  {\bibinfo  {journal} {Phys. Rev. Lett.}\ }\textbf {\bibinfo {volume} {48}},\
  \bibinfo {pages} {344} (\bibinfo {year} {1982})}\BibitemShut {NoStop}%
\bibitem [{\citenamefont {Igl\'oi}\ and\ \citenamefont
  {Monthus}(2005)}]{igloi-review}%
  \BibitemOpen
  \bibfield  {author} {\bibinfo {author} {\bibfnamefont {F.}~\bibnamefont
  {Igl\'oi}}\ and\ \bibinfo {author} {\bibfnamefont {C.}~\bibnamefont
  {Monthus}},\ }\href {\doibase 10.1016/j.physrep.2005.02.006} {\bibfield
  {journal} {\bibinfo  {journal} {Phys. Rep.}\ }\textbf {\bibinfo {volume}
  {412}},\ \bibinfo {pages} {277} (\bibinfo {year} {2005})}\BibitemShut
  {NoStop}%
\bibitem [{\citenamefont {Boechat}\ \emph {et~al.}(1996)\citenamefont
  {Boechat}, \citenamefont {Saguia},\ and\ \citenamefont
  {Continentino}}]{boechatbiquadr}%
  \BibitemOpen
  \bibfield  {author} {\bibinfo {author} {\bibfnamefont {B.}~\bibnamefont
  {Boechat}}, \bibinfo {author} {\bibfnamefont {A.}~\bibnamefont {Saguia}}, \
  and\ \bibinfo {author} {\bibfnamefont {M.~A.}\ \bibnamefont {Continentino}},\
  }\href {\doibase 10.1016/0038-1098(96)00082-8} {\bibfield  {journal}
  {\bibinfo  {journal} {Solid State Commun.}\ }\textbf {\bibinfo {volume}
  {98}},\ \bibinfo {pages} {411} (\bibinfo {year} {1996})}\BibitemShut
  {NoStop}%
\bibitem [{\citenamefont {Hyman}\ and\ \citenamefont
  {Yang}(1997)}]{hymanyang97}%
  \BibitemOpen
  \bibfield  {author} {\bibinfo {author} {\bibfnamefont {R.~A.}\ \bibnamefont
  {Hyman}}\ and\ \bibinfo {author} {\bibfnamefont {K.}~\bibnamefont {Yang}},\
  }\href {\doibase 10.1103/PhysRevLett.78.1783} {\bibfield  {journal} {\bibinfo
   {journal} {Phys. Rev. Lett.}\ }\textbf {\bibinfo {volume} {78}},\ \bibinfo
  {pages} {1783} (\bibinfo {year} {1997})}\BibitemShut {NoStop}%
\bibitem [{\citenamefont {Monthus}\ \emph {et~al.}(1997)\citenamefont
  {Monthus}, \citenamefont {Golinelli},\ and\ \citenamefont
  {Jolicoeur}}]{monthusetal97}%
  \BibitemOpen
  \bibfield  {author} {\bibinfo {author} {\bibfnamefont {C.}~\bibnamefont
  {Monthus}}, \bibinfo {author} {\bibfnamefont {O.}~\bibnamefont {Golinelli}},
  \ and\ \bibinfo {author} {\bibfnamefont {Th.}~\bibnamefont {Jolicoeur}},\
  }\href {\doibase 10.1103/PhysRevLett.79.3254} {\bibfield  {journal} {\bibinfo
   {journal} {Phys. Rev. Lett.}\ }\textbf {\bibinfo {volume} {79}},\ \bibinfo
  {pages} {3254} (\bibinfo {year} {1997})}\BibitemShut {NoStop}%
\bibitem [{\citenamefont {{C. Monthus}}\ \emph {et~al.}(1998)\citenamefont {{C.
  Monthus}}, \citenamefont {{O. Golinelli}},\ and\ \citenamefont {{Th.
  Jolicoeur}}}]{monthusetal98}%
  \BibitemOpen
  \bibfield  {author} {\bibinfo {author} {\bibnamefont {{C. Monthus}}},
  \bibinfo {author} {\bibnamefont {{O. Golinelli}}}, \ and\ \bibinfo {author}
  {\bibnamefont {{Th. Jolicoeur}}},\ }\href {\doibase 10.1103/PhysRevB.58.805}
  {\bibfield  {journal} {\bibinfo  {journal} {Phys. Rev. B}\ }\textbf {\bibinfo
  {volume} {58}},\ \bibinfo {pages} {805} (\bibinfo {year} {1998})}\BibitemShut
  {NoStop}%
\bibitem [{\citenamefont {Saguia}\ \emph {et~al.}(2002)\citenamefont {Saguia},
  \citenamefont {Boechat},\ and\ \citenamefont
  {Continentino}}]{PhysRevLett.89.117202}%
  \BibitemOpen
  \bibfield  {author} {\bibinfo {author} {\bibfnamefont {A.}~\bibnamefont
  {Saguia}}, \bibinfo {author} {\bibfnamefont {B.}~\bibnamefont {Boechat}}, \
  and\ \bibinfo {author} {\bibfnamefont {M.~A.}\ \bibnamefont {Continentino}},\
  }\href {\doibase 10.1103/PhysRevLett.89.117202} {\bibfield  {journal}
  {\bibinfo  {journal} {Phys. Rev. Lett.}\ }\textbf {\bibinfo {volume} {89}},\
  \bibinfo {pages} {117202} (\bibinfo {year} {2002})}\BibitemShut {NoStop}%
\bibitem [{\citenamefont {Bergkvist}\ \emph {et~al.}(2002)\citenamefont
  {Bergkvist}, \citenamefont {Henelius},\ and\ \citenamefont
  {Rosengren}}]{bergekvist-etal-prb02}%
  \BibitemOpen
  \bibfield  {author} {\bibinfo {author} {\bibfnamefont {S.}~\bibnamefont
  {Bergkvist}}, \bibinfo {author} {\bibfnamefont {P.}~\bibnamefont {Henelius}},
  \ and\ \bibinfo {author} {\bibfnamefont {A.}~\bibnamefont {Rosengren}},\
  }\href {\doibase 10.1103/PhysRevB.66.134407} {\bibfield  {journal} {\bibinfo
  {journal} {Phys. Rev. B}\ }\textbf {\bibinfo {volume} {66}},\ \bibinfo
  {pages} {134407} (\bibinfo {year} {2002})}\BibitemShut {NoStop}%
\bibitem [{\citenamefont {Damle}(2002)}]{PhysRevB.66.104425}%
  \BibitemOpen
  \bibfield  {author} {\bibinfo {author} {\bibfnamefont {K.}~\bibnamefont
  {Damle}},\ }\href {\doibase 10.1103/PhysRevB.66.104425} {\bibfield  {journal}
  {\bibinfo  {journal} {Phys. Rev. B}\ }\textbf {\bibinfo {volume} {66}},\
  \bibinfo {pages} {104425} (\bibinfo {year} {2002})}\BibitemShut {NoStop}%
\bibitem [{\citenamefont {Fisher}(1994)}]{fisher94-xxz}%
  \BibitemOpen
  \bibfield  {author} {\bibinfo {author} {\bibfnamefont {D.~S.}\ \bibnamefont
  {Fisher}},\ }\href {\doibase 10.1103/PhysRevB.50.3799} {\bibfield  {journal}
  {\bibinfo  {journal} {Phys. Rev. B}\ }\textbf {\bibinfo {volume} {50}},\
  \bibinfo {pages} {3799} (\bibinfo {year} {1994})}\BibitemShut {NoStop}%
\bibitem [{\citenamefont {Manmana}\ \emph {et~al.}(2011)\citenamefont
  {Manmana}, \citenamefont {L\"auchli}, \citenamefont {Essler},\ and\
  \citenamefont {Mila}}]{manmanaetal11}%
  \BibitemOpen
  \bibfield  {author} {\bibinfo {author} {\bibfnamefont {S.~R.}\ \bibnamefont
  {Manmana}}, \bibinfo {author} {\bibfnamefont {A.~M.}\ \bibnamefont
  {L\"auchli}}, \bibinfo {author} {\bibfnamefont {F.~H.~L.}\ \bibnamefont
  {Essler}}, \ and\ \bibinfo {author} {\bibfnamefont {F.}~\bibnamefont
  {Mila}},\ }\href {\doibase 10.1103/PhysRevB.83.184433} {\bibfield  {journal}
  {\bibinfo  {journal} {Phys. Rev. B}\ }\textbf {\bibinfo {volume} {83}},\
  \bibinfo {pages} {184433} (\bibinfo {year} {2011})}\BibitemShut {NoStop}%
\bibitem [{Sup()}]{Suppl}%
  \BibitemOpen
  \href@noop {} {}\bibinfo {note} {See Supplemental Material, which includes Refs. \cite{K.Yang1996,rodrigopriv,hyman-dimer,PhysRevLett.59.799,chenetal12,hoyos08}, for the relation between spin-1 SU(2) and antisymmetric SU(3) representations, additional numerical data, and discussions of the weak-disorder limit. }\BibitemShut {Stop}%
  
\bibitem [{\citenamefont {{K. Yang}}\ \emph {et~al.}(1996)\citenamefont {{K.
  Yang}}, \citenamefont {{R. A. Hyman}}, \citenamefont {{R. N. Bhatt}},\ and\
  \citenamefont {{S. M. Girvin}}}]{K.Yang1996}%
  \BibitemOpen
  \bibfield  {author} {\bibinfo {author} {\bibnamefont {{K. Yang}}}, \bibinfo
  {author} {\bibnamefont {{R. A. Hyman}}}, \bibinfo {author} {\bibnamefont {{R.
  N. Bhatt}}}, \ and\ \bibinfo {author} {\bibnamefont {{S. M. Girvin}}},\
  }\href {\doibase 10.1063/1.361312} {\bibfield  {journal} {\bibinfo  {journal}
  {J. Appl. Phys.}\ }\textbf {\bibinfo {volume} {79}},\ \bibinfo {pages} {5096}
  (\bibinfo {year} {1996})}\BibitemShut {NoStop}%
\bibitem [{\citenamefont {Pereira}()}]{rodrigopriv}%
  \BibitemOpen
  \bibfield  {author} {\bibinfo {author} {\bibfnamefont {R.~G.}\ \bibnamefont
  {Pereira}},\ }\href@noop {} {}\bibinfo {note} {(private
  communication)}\BibitemShut {NoStop}%
\bibitem [{\citenamefont {Hyman}\ \emph {et~al.}(1996)\citenamefont {Hyman},
  \citenamefont {Yang}, \citenamefont {Bhatt},\ and\ \citenamefont
  {Girvin}}]{hyman-dimer}%
  \BibitemOpen
  \bibfield  {author} {\bibinfo {author} {\bibfnamefont {R.~A.}\ \bibnamefont
  {Hyman}}, \bibinfo {author} {\bibfnamefont {K.}~\bibnamefont {Yang}},
  \bibinfo {author} {\bibfnamefont {R.~N.}\ \bibnamefont {Bhatt}}, \ and\
  \bibinfo {author} {\bibfnamefont {S.~M.}\ \bibnamefont {Girvin}},\ }\href
  {\doibase 10.1103/PhysRevLett.76.839} {\bibfield  {journal} {\bibinfo
  {journal} {Phys. Rev. Lett.}\ }\textbf {\bibinfo {volume} {76}},\ \bibinfo
  {pages} {839} (\bibinfo {year} {1996})}\BibitemShut {NoStop}%
\bibitem [{\citenamefont {Affleck}\ \emph {et~al.}(1987)\citenamefont
  {Affleck}, \citenamefont {Kennedy}, \citenamefont {Lieb},\ and\ \citenamefont
  {Tasaki}}]{PhysRevLett.59.799}%
  \BibitemOpen
  \bibfield  {author} {\bibinfo {author} {\bibfnamefont {I.}~\bibnamefont
  {Affleck}}, \bibinfo {author} {\bibfnamefont {T.}~\bibnamefont {Kennedy}},
  \bibinfo {author} {\bibfnamefont {E.~H.}\ \bibnamefont {Lieb}}, \ and\
  \bibinfo {author} {\bibfnamefont {H.}~\bibnamefont {Tasaki}},\ }\href
  {\doibase 10.1103/PhysRevLett.59.799} {\bibfield  {journal} {\bibinfo
  {journal} {Phys. Rev. Lett.}\ }\textbf {\bibinfo {volume} {59}},\ \bibinfo
  {pages} {799} (\bibinfo {year} {1987})}\BibitemShut {NoStop}%
\bibitem [{\citenamefont {Chen}\ \emph {et~al.}(2012)\citenamefont {Chen},
  \citenamefont {Gu}, \citenamefont {Liu},\ and\ \citenamefont
  {Wen}}]{chenetal12}%
  \BibitemOpen
  \bibfield  {author} {\bibinfo {author} {\bibfnamefont {X.}~\bibnamefont
  {Chen}}, \bibinfo {author} {\bibfnamefont {Z.-C.}\ \bibnamefont {Gu}},
  \bibinfo {author} {\bibfnamefont {Z.-X.}\ \bibnamefont {Liu}}, \ and\
  \bibinfo {author} {\bibfnamefont {X.-G.}\ \bibnamefont {Wen}},\ }\href
  {\doibase 10.1126/science.1227224} {\bibfield  {journal} {\bibinfo  {journal}
  {Science}\ }\textbf {\bibinfo {volume} {338}},\ \bibinfo {pages} {1604}
  (\bibinfo {year} {2012})}\BibitemShut {NoStop}%
\bibitem [{\citenamefont {Hoyos}(2008)}]{hoyos08}%
  \BibitemOpen
  \bibfield  {author} {\bibinfo {author} {\bibfnamefont {J.~A.}\ \bibnamefont
  {Hoyos}},\ }\href {\doibase 10.1103/PhysRevE.78.032101} {\bibfield  {journal}
  {\bibinfo  {journal} {Phys. Rev. E}\ }\textbf {\bibinfo {volume} {78}},\
  \bibinfo {pages} {032101} (\bibinfo {year} {2008})}\BibitemShut {NoStop}%
\bibitem [{Note1()}]{Note1}%
  \BibitemOpen
  \bibinfo {note} {$K_{i}$ is a natural coupling constant when \protect \textup
  {\hbox {\mathsurround \z@ \protect \normalfont (\ignorespaces \ref
  {eq:hamilt}\unskip \@@italiccorr )}} is written in terms of irreducible
  spherical tensors, as noted in \cite {PhysRevLett.80.4562}}\BibitemShut
  {NoStop}%
\bibitem [{\citenamefont {Yang}\ and\ \citenamefont
  {Bhatt}(1998)}]{PhysRevLett.80.4562}%
  \BibitemOpen
  \bibfield  {author} {\bibinfo {author} {\bibfnamefont {K.}~\bibnamefont
  {Yang}}\ and\ \bibinfo {author} {\bibfnamefont {R.~N.}\ \bibnamefont
  {Bhatt}},\ }\href {\doibase 10.1103/PhysRevLett.80.4562} {\bibfield
  {journal} {\bibinfo  {journal} {Phys. Rev. Lett.}\ }\textbf {\bibinfo
  {volume} {80}},\ \bibinfo {pages} {4562} (\bibinfo {year}
  {1998})}\BibitemShut {NoStop}%
\bibitem [{\citenamefont {Hoyos}\ and\ \citenamefont
  {Miranda}(2004)}]{PhysRevB.70.180401}%
  \BibitemOpen
  \bibfield  {author} {\bibinfo {author} {\bibfnamefont {J.~A.}\ \bibnamefont
  {Hoyos}}\ and\ \bibinfo {author} {\bibfnamefont {E.}~\bibnamefont
  {Miranda}},\ }\href {\doibase 10.1103/PhysRevB.70.180401} {\bibfield
  {journal} {\bibinfo  {journal} {Phys. Rev. B}\ }\textbf {\bibinfo {volume}
  {70}},\ \bibinfo {pages} {180401} (\bibinfo {year} {2004})}\BibitemShut
  {NoStop}%
\bibitem [{\citenamefont {{E. Westerberg}}\ \emph {et~al.}(1997)\citenamefont
  {{E. Westerberg}}, \citenamefont {{A. Furusaki}}, \citenamefont {{M.
  Sigrist}},\ and\ \citenamefont {{P. A. Lee}}}]{westerbergetal}%
  \BibitemOpen
  \bibfield  {author} {\bibinfo {author} {\bibnamefont {{E. Westerberg}}},
  \bibinfo {author} {\bibnamefont {{A. Furusaki}}}, \bibinfo {author}
  {\bibnamefont {{M. Sigrist}}}, \ and\ \bibinfo {author} {\bibnamefont {{P. A.
  Lee}}},\ }\href {\doibase 10.1103/PhysRevB.55.12578} {\bibfield  {journal}
  {\bibinfo  {journal} {Phys. Rev. B}\ }\textbf {\bibinfo {volume} {55}},\
  \bibinfo {pages} {12578} (\bibinfo {year} {1997})}\BibitemShut {NoStop}%
\bibitem [{\citenamefont {{V. L. Quito}}\ \emph {et~al.}()\citenamefont {{V. L.
  Quito}}, \citenamefont {{Jos\'e A. Hoyos}},\ and\ \citenamefont {{E.
  Miranda}}}]{quito-hoyos-miranda}%
  \BibitemOpen
  \bibfield  {author} {\bibinfo {author} {\bibnamefont {{V. L. Quito}}},
  \bibinfo {author} {\bibnamefont {{J. A. Hoyos}}}, \ and\ \bibinfo
  {author} {\bibnamefont {{E. Miranda}}},\ }\href@noop {} {}\bibinfo {note}
  {(unpublished)}\BibitemShut {NoStop}%
\bibitem [{Note2()}]{Note2}%
  \BibitemOpen
  \bibinfo {note} {In general, singlets of 6, 9,... original spins/quarks are
  also formed, though less abundantly \cite {PhysRevB.70.180401}.}\BibitemShut
  {Stop}%
\bibitem [{\citenamefont {Hoyos}\ \emph {et~al.}(2007)\citenamefont {Hoyos},
  \citenamefont {Vieira}, \citenamefont {Laflorencie},\ and\ \citenamefont
  {Miranda}}]{PhysRevB.76.174425}%
  \BibitemOpen
  \bibfield  {author} {\bibinfo {author} {\bibfnamefont {J.~A.}\ \bibnamefont
  {Hoyos}}, \bibinfo {author} {\bibfnamefont {A.~P.}\ \bibnamefont {Vieira}},
  \bibinfo {author} {\bibfnamefont {N.}~\bibnamefont {Laflorencie}}, \ and\
  \bibinfo {author} {\bibfnamefont {E.}~\bibnamefont {Miranda}},\ }\href
  {\doibase 10.1103/PhysRevB.76.174425} {\bibfield  {journal} {\bibinfo
  {journal} {Phys. Rev. B}\ }\textbf {\bibinfo {volume} {76}},\ \bibinfo
  {pages} {174425} (\bibinfo {year} {2007})}\BibitemShut {NoStop}%
\end{thebibliography}
\end{document}


\title{Supplemental Material for ``Emergent SU(3) symmetry in random spin-1
chains''}

\author{V. L. Quito}

\affiliation{Instituto de F\'{\i}sica Gleb Wataghin, Unicamp, Rua S\'{e}rgio Buarque de
Holanda, 777, CEP 13083-859 Campinas, SP, Brazil}

\author{Jos\'{e} A. Hoyos}

\affiliation{Instituto de F\'{\i}sica de S\~{a}o Carlos, Universidade de S\~{a}o Paulo, C.P.
369, S\~{a}o Carlos, SP 13560-970, Brazil}

\author{E. Miranda}

\affiliation{Instituto de F\'{\i}sica Gleb Wataghin, Unicamp, Rua S\'{e}rgio Buarque de
Holanda, 777, CEP 13083-859 Campinas, SP, Brazil}

\selectlanguage{english}%

\date{\today}

\maketitle
\selectlanguage{american}%

\section{The SU(3)-symmetric points}

We first write out in detail the 8 generators of the SU(3) group in
the defining (quark) representation in terms of the spin-1 operators.
The Cartesian vector and symmetric rank-2 tensor operators, $S_{\mu}$
and $T_{\mu\nu}=S_{\mu}S_{\nu}+S_{\nu}S_{\mu}$ ($\mu,\nu=x,y,z$),
form a complete basis set of 9 elements spanning the space of $3\times3$
Hermitian matrices. The generators we seek are the complete set of
8 \emph{traceless} Hermitian matrices. A convenient choice is
\begin{eqnarray}
\Lambda_{1} & = & S_{x},\label{eq:sx}\\
\Lambda_{2} & = & S_{y},\label{eq:sy}\\
\Lambda_{3} & = & S_{z},\\
\Lambda_{4} & = & S_{x}S_{y}+S_{y}S_{x},\label{eq:qxy}\\
\Lambda_{5} & = & S_{x}S_{z}+S_{z}S_{x},\label{eq:qxz}\\
\Lambda_{6} & = & S_{y}S_{z}+S_{z}S_{y},\label{eq:qyz}\\
\Lambda_{7} & = & S_{x}^{2}-S_{y}^{2},\label{eq:qx2y2}\\
\Lambda_{8} & = & \frac{1}{\sqrt{3}}\left(2S_{z}^{2}-S_{x}^{2}-S_{y}^{2}\right).\label{eq:q0}
\end{eqnarray}

In terms of these generators, the linear Heisenberg term is obvious,
whereas the bilinear term can be written as
\begin{equation}
\left(\mathbf{S}\cdot\mathbf{S}^{\prime}\right)^{2}=\frac{4}{3}-\frac{1}{2}\sum_{a=1}^{3}\Lambda_{a}\Lambda_{a}^{\prime}+\frac{1}{2}\sum_{a=4}^{8}\Lambda_{a}\Lambda_{a}^{\prime}.
\end{equation}
The generic Hamiltonian for two adjacent sites in terms of $\theta$
($\tan\theta=D/J$) is
\begin{eqnarray}
H\left(\theta\right) & = & \cos\theta\mathbf{S}\cdot\mathbf{S}^{\prime}+\sin\theta\left(\mathbf{S}\cdot\mathbf{S}^{\prime}\right)^{2}\label{eq:hamtheta}\\
 & = & \frac{4}{3}\sin\theta+\left(\cos\theta-\frac{1}{2}\sin\theta\right)\sum_{a=1}^{3}\Lambda_{a}\Lambda_{a}^{\prime}\\
 &  & +\frac{1}{2}\sin\theta\sum_{a=4}^{8}\Lambda_{a}\Lambda_{a}^{\prime}.
\end{eqnarray}
The point $\theta=\frac{\pi}{4}$ then becomes
\begin{eqnarray}
H\left(\theta=\frac{\pi}{4}\right) & \propto & \mathbf{S}\cdot\mathbf{S}^{\prime}+\left(\mathbf{S}\cdot\mathbf{S}^{\prime}\right)^{2}\\
 & = & \frac{4}{3}+\frac{1}{2}\sum_{a=1}^{8}\Lambda_{a}\Lambda_{a}^{\prime},
\end{eqnarray}
whose SU(3) invariance is manifest. The other SU(3)-invariant point
at $\theta=-\frac{\pi}{2}$ is 
\begin{eqnarray}
H\left(\theta=-\frac{\pi}{2}\right) & = & -\left(\mathbf{S}\cdot\mathbf{S}^{\prime}\right)^{2}\\
 & = & -\frac{4}{3}+\frac{1}{2}\sum_{a=1}^{3}\Lambda_{a}\Lambda_{a}^{\prime}-\frac{1}{2}\sum_{a=4}^{8}\Lambda_{a}\Lambda_{a}^{\prime}.\label{eq:minuspiover2}
\end{eqnarray}
In order to make the SU(3) invariance manifest in this case, we must
first realize that the antiquark representation, the complex conjugate
of the quark one, is obtained by applying the following operation
to the generator matrices
\begin{equation}
\left(\Lambda_{a}\right)_{ij}\to-\left(\Lambda_{a}\right)_{ij}^{\ast}.
\end{equation}
We get (assuming the usual basis of eigenvectors of $S_{z}$)
\begin{eqnarray}
\Lambda_{1} & \to & -S_{x},\label{eq:sx-1}\\
\Lambda_{2} & \to & S_{y},\label{eq:sy-1}\\
\Lambda_{3} & \to & -S_{z},\\
\Lambda_{4} & \to & S_{x}S_{y}+S_{y}S_{x},\label{eq:qxy-1}\\
\Lambda_{5} & \to & -S_{x}S_{z}-S_{z}S_{x},\label{eq:qxz-1}\\
\Lambda_{6} & \to & S_{y}S_{z}+S_{z}S_{y},\label{eq:qyz-1}\\
\Lambda_{7} & \to & -S_{x}^{2}+S_{y}^{2},\label{eq:qx2y2-1}\\
\Lambda_{8} & \to & -\frac{1}{\sqrt{3}}\left(2S_{z}^{2}-S_{x}^{2}-S_{y}^{2}\right).\label{eq:q0-1}
\end{eqnarray}
An equivalent representation to Eqs.~(\ref{eq:sx-1})-(\ref{eq:q0-1})
is obtained if we make a rotation of $\pi$ around the $y$-axis:
$S_{x}\to-S_{x}$, $S_{z}\to-S_{z}$. After these transformations,
we find, denoting the antiquark representation by a tilde,
\begin{eqnarray}
\tilde{\Lambda}_{a} & = & \Lambda_{a}\,\left(a=1,2,3\right),\label{eq:antiquark1}\\
\tilde{\Lambda}_{a} & = & -\Lambda_{a}\,\left(a=4,5,6,7,8\right).\label{eq:antiquark2}
\end{eqnarray}
From Eqs.~\eqref{eq:minuspiover2} and (\ref{eq:antiquark1})-(\ref{eq:antiquark2}),
we see that the $\theta=-\frac{\pi}{2}$ point couples sites belonging
to the quark and the antiquark representations 
\begin{equation}
H\left(\theta=-\frac{\pi}{2}\right)=-\frac{4}{3}+\frac{1}{2}\sum_{a=1}^{8}\Lambda_{a}\tilde{\Lambda}_{a}^{\prime}.
\end{equation}
Note that the FM points $\theta=-\frac{3\pi}{4}$ and $\theta=\frac{\pi}{2}$
also display global SU(3) invariance since
\begin{eqnarray}
H\left(\theta=\frac{\pi}{2}\right) & = & -H\left(\theta=-\frac{\pi}{2}\right),\\
H\left(\theta=\frac{3\pi}{4}\right) & = & -H\left(\theta=\frac{\pi}{4}\right).
\end{eqnarray}

\section{Numerical results}

We have implemented the full SDRG procedure described in the main
text numerically in order to check our findings. We have first focused
on initial Hamiltonians in which all bonds have the same $\theta_{i}=\theta_{0}$
with $-\frac{3\pi}{4}<\theta_{0}<\frac{\pi}{2}$, while $J$ is uniformly
distributed in the interval $0\le J\le1$. The results were obtained
for chain lengths of $L_{0}\sim10^{6}$ spins, averaged over 20 realizations
of disorder. We have verified that, although initially the $\theta$
distribution broadens, \emph{asymptotically all $\theta_{i}$ tend
to unique values}. 

\begin{figure}[h]
\begin{centering}
\includegraphics[clip,scale=0.7]{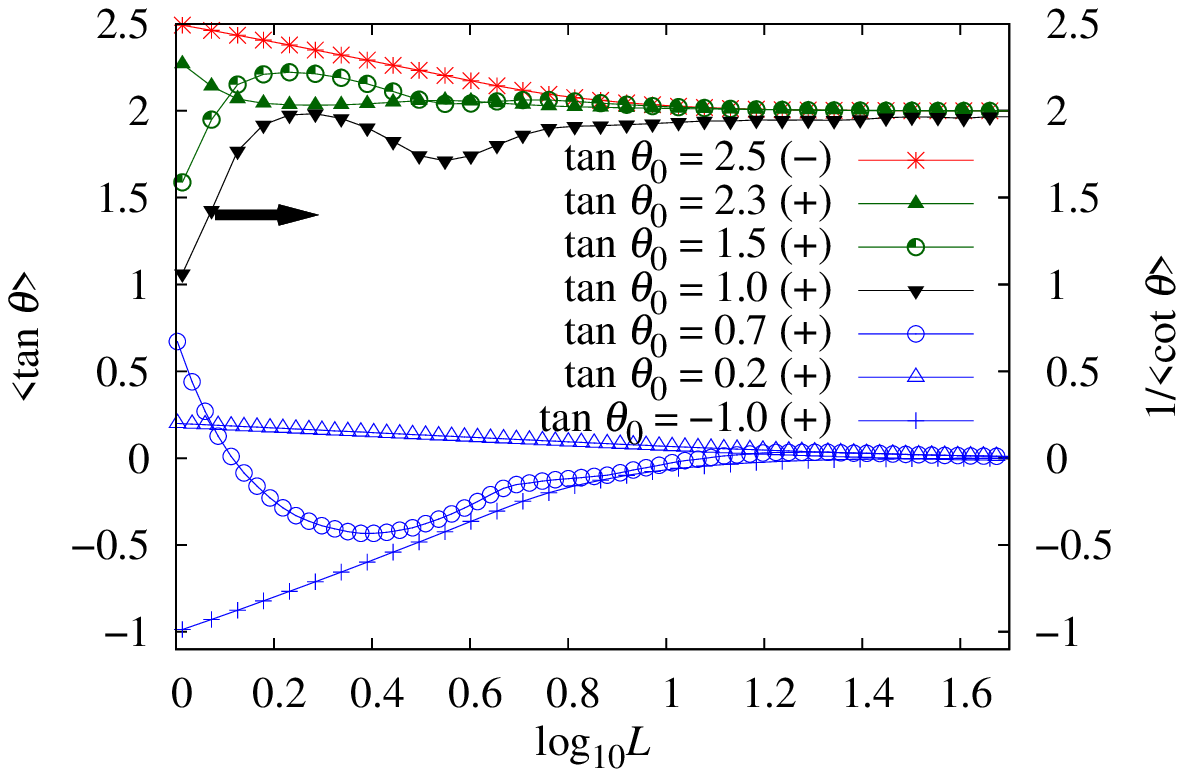}
\par\end{centering}

\caption{\label{fig:Mean-values} (Color online) Some representative SDRG flows:
mean value of $\left\langle \tan\theta\right\rangle $ as a function
of the average distance $L$ between undecimated spin clusters, for
different values of the angle $\theta_{0}$ of the initial distribution.
The sign in parentheses is the sign of $J$'s in the initial distribution.
For $\tan\theta_{0}=1$, we plot $\frac{1}{\left\langle \cot\theta\right\rangle }$
on the right-hand vertical axis instead.}
\end{figure}

In Fig.~\ref{fig:Mean-values}, we show, for some representative
cases, the average value of $\tan\theta$ as the mean distance $L$
between undecimated spin clusters is increased. The latter is given
by $L=\frac{L_{0}}{N}$, where $N$ is the number of undecimated spin
clusters. For $-\frac{\pi}{2}<\theta_{0}<\frac{\pi}{4}$ (three blue
lowest curves), the flow is towards the FP (1), with $\theta=0$.
When $-\frac{3\pi}{4}<\theta_{0}<-\frac{\pi}{2}$ (red, topmost curve),
the flow tends to the FP (2), with $\tan\theta=2$, $D<0$. The green
(2nd and 3rd from the top) curves correspond to the interval $\frac{\pi}{4}<\theta_{0}<\frac{\pi}{2}$,
for which the systems flows to the FP (3), $\tan\theta=2$, $D\lessgtr0$
with equal probability. For the specific case of $\theta_{0}=\frac{\pi}{4}$
(black curve with down triangles), we plotted instead the inverse
of $\left\langle \cot\theta\right\rangle $ on the right-hand vertical
axis. This corresponds to flows at the unstable FP (8), for which
asymptotically half the bonds have $\theta=-\frac{\pi}{2}$ and half
$\theta=\frac{\pi}{4}$, such that $\left\langle \cot\theta\right\rangle \to1/2$.
These plots confirm the delineation of the fixed points and basins
of attraction described in the main text.

Although the flow in the region ($-\frac{\pi}{2}<\theta_{0}<\frac{\pi}{4}$)
is characterized by the disappearance of the biquadratic couplings,
there is a clear difference \textcolor{black}{in the transient SDRG
flow }between the cases $\arctan\frac{1}{3}<\theta_{0}<\frac{\pi}{4}$
and $-\frac{\pi}{2}<\theta_{0}<\arctan\frac{1}{3}$. The latter only
involves singlet-forming decimations {[}see Eq.~(2) and Fig.~2 of
the main text{]}. As a result, $\left\langle \tan\theta\right\rangle $
flows monotonically to 0 ($\tan\theta_{0}=0.2$ and $-1$ in Fig.~\ref{fig:Mean-values}).
When $\arctan\frac{1}{3}<\theta_{0}<\frac{\pi}{4}$, however, both
types of decimations rules occur {[}Eqs.~\textcolor{black}{(2) and
(3)} of the main text{]}, necessarily generating negative angles,
in such a way that $\left\langle \tan\theta\right\rangle $ may change
sign in the course of the flow ($\tan\theta_{0}=0.7$ in Fig.~\ref{fig:Mean-values}). 

\begin{figure}[h]
\begin{centering}
\includegraphics[clip,scale=0.65]{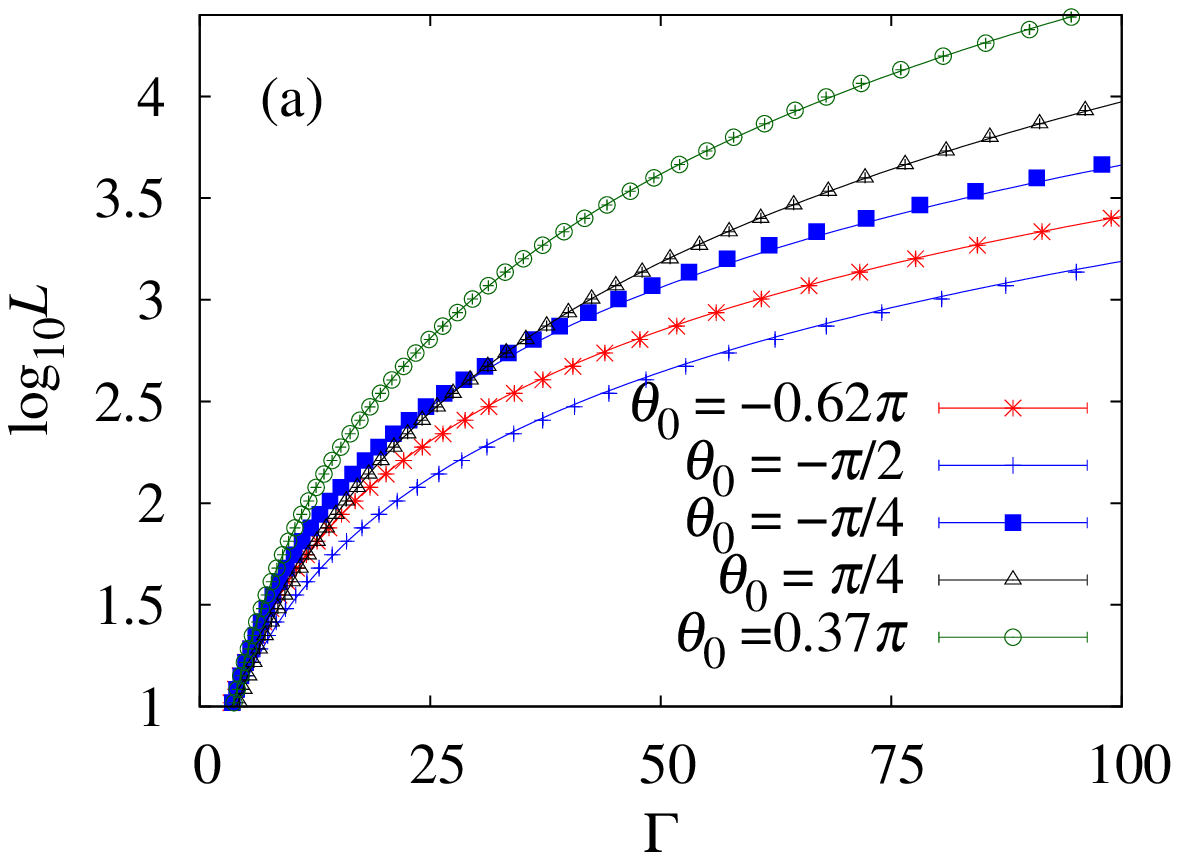}
\includegraphics[clip,scale=0.65]{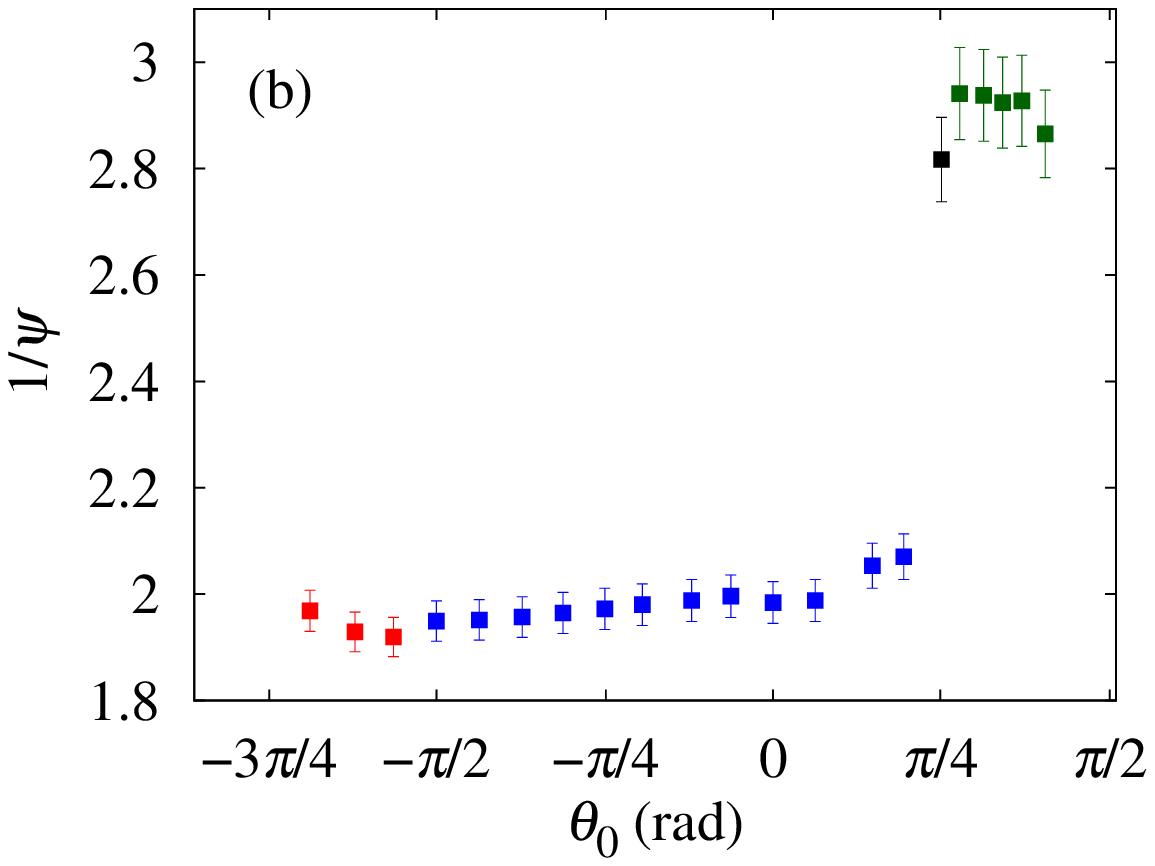}
\par\end{centering}

\caption{\label{fig:din_exponent}(Color online) Numerical determination of
the $\psi$ exponent: (a) mean distance $L$ between undecimated spin
clusters as a function of $\Gamma=\ln\frac{\Omega_{0}}{\Omega}$,
where $\Omega$ is the renormalization group energy scale. The full
lines are fits to Eq.~\eqref{eq:fitting}; (b) reciprocal of the
activated dynamical scaling exponent $\psi$ as a function of initial
angle $\theta_{0}$. }
\end{figure}

In order to numerically determine the value of $\psi$ in each of
the RSPs of the phase diagram, we tracked the dependence of the cutoff
\textcolor{black}{energy} scale $\Gamma\equiv\ln\frac{\Omega_{0}}{\Omega}$
(where $\Omega_{0}$ is the largest initial $\Delta_{i}$) on the
average distance between undecimated spin clusters $L$. In Fig.~\ref{fig:din_exponent}(a)
we show the results for several initial angles $\theta_{0}$. We have
fitted our data to the form

\begin{equation}
\log_{10}L=a+\frac{1}{\psi}\log_{10}\left(1+b\Gamma\right),\label{eq:fitting}
\end{equation}
where $a,\, b$ and $\frac{1}{\psi}$ are fitting parameters. As seen
in Fig.~\ref{fig:din_exponent}(a), it fits remarkably well the numerical
data. For $b\Gamma\gg1$, we recover the more familiar activated dynamical
scaling form $L\sim\Gamma^{\frac{1}{\psi}}$. The numerically determined
values of $\psi$ can be seen in Fig.~\ref{fig:din_exponent}(b)
as a function of $\theta_{0}$. There is good agreement with the predicted
exponents of the mesonic ($\psi_{M}=\frac{1}{2}$) and baryonic ($\psi_{B}=\frac{1}{3}$)
RSPs and a sharp jump at the border between them ($\theta_{0}=\frac{\pi}{4}$)
can be clearly seen. The error bars were estimated from the uncertainty
in the range over which Eq.~\ref{eq:fitting} is valid.

\section{Behavior at weak disorder}

The spontaneously dimerized phase ($-\frac{3\pi}{4}<\theta<-\frac{\pi}{4}$)
is unstable against weak disorder due to the formation of weakly coupled
domain walls~\cite{K.Yang1996,boechatbiquadr}. Weak disorder can
also be shown to be a relevant perturbation at the SU(3)-symmetric
point $\theta=\frac{\pi}{4}$ \cite{rodrigopriv}. In general, we
expect disorder to be perturbatively relevant in the entire gapless
phase $\frac{\pi}{4}\leq\theta\leq\frac{\pi}{2}$. 

Infinitesimally weak disorder is an irrelevant perturbation in the
gapped Haldane phase ($-\frac{\pi}{4}<\theta<\frac{\pi}{4}$). The
behavior was determined in detail at $\theta_{i}=0$~\cite{boechatbiquadr,hymanyang97,monthusetal97,monthusetal98,PhysRevLett.89.117202}.
Gradually increasing the disorder at this point eventually leads to
the closure of the Haldane gap (although the topological order parameter
initially retains a finite value~\cite{hyman-dimer}) and to the
emergence of a quantum Griffiths region~\cite{PhysRevB.66.104425}
with conventional power-law scaling $\Omega\sim L^{-z}$. In this
region, the spin correlations are short ranged, the magnetic susceptibility
$\chi\sim T^{1/z-1}$, and the specific heat $c\sim T^{1/z}$. The
dynamical exponent $z$ is disorder-dependent and diverges at a critical
point~\cite{hymanyang97,monthusetal97,monthusetal98,PhysRevLett.89.117202,bergekvist-etal-prb02,PhysRevB.66.104425}.
Above this critical disorder value, the system enters a universal
RSP governed by an IRFP with $\psi=\frac{1}{2}$. This generic behavior
is expected to hold throughout the region $-\frac{\pi}{4}<\theta<\frac{\pi}{4}$,
with the exception of the AKLT point.

At the AKLT point ($\theta=\arctan\frac{1}{3}$), the ground state
is exactly known~\cite{PhysRevLett.59.799}. Provided the local angle
$\theta_{i}$ is everywhere equal to $\arctan\frac{1}{3}$, the ground
state will be disorder independent \cite{chenetal12}. We note that
this is reflected in the SDRG procedure by the closure of the gap
of a spin pair $\Delta_{i}=3J_{i}\left|\tan\theta_{i}-\frac{1}{3}\right|$,
which makes it ill-defined in the vicinity of the AKLT point. The
Haldane gap, however, will vanish in the strong-disorder limit when
the distribution of coupling constants is not bounded from below.

Although tailored to be accurate only in the strong disorder regime,
the SDRG flow has been shown to break down whenever weak disorder
is irrelevant \cite{boechatbiquadr,hymanyang97,monthusetal97,monthusetal98,PhysRevLett.89.117202,PhysRevB.66.104425,hoyos08}.
This is signaled by the fact that the coupling constant distributions
do not broaden as the energy scale is reduced. We only detect this
break-down in the topological phase $-\frac{\pi}{4}<\theta<\frac{\pi}{4}$~%
\footnote{In this case, it is still possible to use the SDRG through an appropriate
generalization of the recursion relations as in reference \cite{monthusetal97}.%
}. In fact, our numerics indicate that weak disorder may possibly be
relevant even inside the Haldane phase near the edges $\pm\frac{\pi}{4}$.
Since the dimerized ($-\frac{3\pi}{4}<\theta<-\frac{\pi}{4}$) and
the gapless ($\frac{\pi}{4}<\theta<\frac{\pi}{2}$) phases are expected
to be destabilized by any amount of disorder and we have not found
any other fixed point numerically, we conjecture that no other phase
transition happens at intermediate disorder strength. We thus obtain
the weak disorder region of the phase diagram sketched in Fig.~1
of the main text.

\bibliographystyle{apsrev4-1}

%